\documentclass[
 reprint,
superscriptaddress,
 amsmath,amssymb,
prb,
]{revtex4-1}

\usepackage{color}
\usepackage{graphicx}% Include figure files
\usepackage{dcolumn}% Align table columns on decimal point
\usepackage{bm}% bold math
\usepackage{hyperref}% add hypertext capabilities
\usepackage[mathlines]{lineno}% Enable numbering of text and display math
\usepackage{textcomp}
\usepackage{natbib}
\begin{document}

%\preprint{APS/123-QED}

\title{Domain Wall Orientations in Ferroelectric Superlattices Probed with Synchrotron X-Ray Diffraction}% Force line breaks with \\

\author{Marios Hadjimichael}
 \email{marios.hadjimichael.14@ucl.ac.uk}
\affiliation{%
	London Centre for Nanotechnology and Department of Physics and Astronomy, University
	College London, 17–-19 Gordon Street, WC1H 0AH, London, United Kingdom
}%
\author{Edoardo Zatterin}%

\affiliation{%
 London Centre for Nanotechnology and Department of Physics and Astronomy, University
 College London, 17–-19 Gordon Street, WC1H 0AH, London, United Kingdom
}%
 \affiliation{ESRF –- The European Synchrotron, 71 Avenue des Martyrs, 38000, Grenoble, France}
 
\author{St{\'{e}}phanie Fernandez-Pe{\~{n}}a} 
 \affiliation{Department of Quantum Matter Physics, University of Geneva, CH-1211 Geneva, Switzerland}

\author{Steven J. Leake}
 \affiliation{ESRF –- The European Synchrotron, 71 Avenue des Martyrs, 38000, Grenoble, France}
\author{Pavlo Zubko}%
\affiliation{%
	London Centre for Nanotechnology and Department of Physics and Astronomy, University
	College London, 17–-19 Gordon Street, WC1H 0AH, London, United Kingdom
}%

\date{\today}% It is always \today, today,

\begin{abstract}
Ferroelectric domains in PbTiO$_3$/SrTiO$_3$ superlattices were studied using synchrotron X-ray diffraction. Macroscopic measurements revealed a change in the domain wall orientation from $\left\lbrace 100 \right\rbrace $ to $\left\lbrace 110 \right\rbrace $ crystallographic planes with increasing temperature. The temperature range of this reorientation depends on the ferroelectric layer thickness and domain period. Using a nanofocused beam, local changes in domain wall orientation within the buried ferroelectric layers were imaged, both in structurally uniform regions of the sample and near defect sites and argon ion etched patterns. Domain walls were found to exhibit preferential alignment with the straight edges of the etched patterns as well as with structural features associated with defect sites. The distribution of out-of-plane lattice parameters was mapped around one such feature, showing that it is accompanied by inhomogeneous strain and large strain gradients.

%\begin{description}
%\end{description}
\end{abstract}

\pacs{Valid PACS appear here}% PACS, the Physics and Astronomy
\maketitle

%\tableofcontents
\section{\label{sec:level1}Introduction}

Recent years have witnessed a renewed interest in ferroelectric domains and domain walls (DWs), particularly at the nanoscale, where DW densities are large and their properties contribute greatly to the overall functional behavior of the materials \cite{Catalan2012,Salje2009}. For example, DW motion is known to greatly influence the macroscopic properties of ferroelectrics, such as their response to an external electric field or external stress \cite{Zhang1994} and  recently has been shown to give rise to the phenomenon of negative capacitance \cite{Bratkovsky2000,Zubko2016a}. DWs also exhibit symmetry and hence properties that are different from the bulk material. Striking examples include the conducting DWs in insulating multiferroic BiFeO$_3$ \cite{Seidel2009,Farokhipoor2011} and ferromagnetic DWs in TbMnO$_3$ \cite{Farokhipoor2014}. Such properties make multidomain ferroelectrics an attractive pathway towards novel device implementations being explored by the growing field of DW nanoelectronics \cite{Catalan2012}. Development of macroscopic methods to control the formation and orientation of DWs is thus central to making device fabrication easier using conventional growth and patterning methods.

Ferroelectric superlattices (periodic repetitions of ferroelectric/dielectric bilayers) are ideal systems for the study of ferroelectric domains. The domains in these systems have well-defined periodicities, making them particularly well-suited for X-ray diffraction studies \cite{Streiffer2002}. In the past, such superlattices have been used to investigate electrostatic coupling through dielectric layers \cite{Specht1998, Stephanovich2005, Zubko2012b}, DW motion under applied electric field \cite{Zubko2010,Jo2011}, improper ferroelectricity \cite{Bousquet2008}, and ferroelectric vortices \cite{Yadav2015}.

In this paper, we present a synchrotron X-ray diffraction study of DW orientations in PbTiO$_3$/SrTiO$_3$ superlattices. We look at the behavior of ferroelectric domains in these materials across the ferroelectric phase transition and discover a temperature regime where the DWs change their orientation from walls that lie preferentially within the $\left\lbrace 100 \right\rbrace $ crystallographic planes at room temperature to predominantly $\left\lbrace 110 \right\rbrace $ DWs at higher temperatures, with a temperature range that depends on the ferroelectric layer thickness and domain period. Using a nanofocused beam, we obtain maps of the local orientations of the buried ferroelectric domains. Imaging the domain orientations near the edge of the superlattice and in the vicinity of a microscopic defect reveals that DWs tend to align preferentially along the film edges and features associated with the structure of the defect. Furthermore, we obtain local three-dimensional maps of reciprocal space at different points around one of these features and use them to reconstruct images of the local lattice parameter, strain and strain gradient distributions that are likely to be responsible for the observed DW alignments. Our findings demonstrate that the orientation of 180$^{\circ}$ DWs can be controlled by varying temperature, film thickness and topographical features of the samples, and suggest that patterning features by conventional photolithography or focused ion beam techniques could offer additional new pathways for manipulating DWs for nanodevices \cite{Whyte2014,Whyte2015}.

\section{\label{sec:level2}Experimental methods}

PbTiO$_3$/SrTiO$_3$ superlattices were grown on SrTiO$_3$ substrates using off-axis radiofrequency magnetron sputtering in a 0.18 Torr oxygen/argon atmosphere of ratio 5:7 with a substrate temperature of 520$^{\circ}$C. Top and bottom epitaxial SrRuO$_3$ electrodes with a thickness of approximately 30 nm were deposited in 0.1 Torr of 1:20 oxygen/argon atmosphere and at a substrate temperature of 650$^{\circ}$C. Here we report the results for two superlattices with compositions (PTO$_{5}$$\vert$STO$_{4}$)$_{28}$ and (PTO$_{8}$$\vert$STO$_{4}$)$_{19}$ where the notation (PTO$_{n_P}$$\vert$STO$_{n_S}$)$_N$ refers to $n_P$ unit cells (u.c.) of PbTiO$_3$ and $n_S$ u.c. of SrTiO$_3$ per period, repeated $N$ times. The (PTO$_{5}$$\vert$STO$_{4}$)$_{28}$  superlattice was deposited on a TiO$_2$-terminated SrTiO$_3$ substrate, whereas the (PTO$_{8}$$\vert$STO$_{4}$)$_{19}$ superlattice was deposited on Nb-doped SrTiO$_3$. The top electrodes were patterned into 120 \textmu{}m $\times$ 120 \textmu{}m squares using UV photolithography and argon ion milling. The (PTO$_{8}$$\vert$STO$_{4}$)$_{19}$ superlattice was etched fully down to the bottom electrode, leaving only the square capacitor structures ($\sim$100 nm thick). Atomic force microscopy (AFM) measurements on the surface of one capacitor revealed a number of micron-sized topographical features that might be regions which were partially etched during argon milling, or other extended structural defects in the substrate and/or film. These features, along with the edges of the square capacitor structures, allowed us to study the impact of topography variations and structural defects on DW orientations. 

X-ray diffraction experiments were performed at beamline ID01 at ESRF. For temperature-dependent measurements, an incoherent X-ray beam (approximately 50 \textmu{}m $\times$ 50 \textmu{}m full-width at half-maximum (FWHM)) was used to probe the average domain behavior. The spatial variation of the domain intensities was studied by focusing the beam with a Fresnel zone plate (achieving a beam size of approximately 60 nm FWHM) with the sample mounted on a resistive heater on a piezo stage. The setup at beamline ID01 has been described in detail in Refs \onlinecite{Chahine2014a,Keplinger2015,Chahine2015,Leake2017}. The energy of the incident X-ray beam was tuned to 8 keV using a Si(111) double crystal monochromator and diffraction patterns were recorded with a Maxipix photon counting detector \cite{Ponchut2011}. The spatial dependence of lattice parameters was calculated using the strain and orientation calculation software package X-SOCS \cite{Chahine2014a,XSOCSgit}. 

\section{\label{sec:level3}Results}
\subsection{\label{sec:temperature}Temperature dependence of ferroelectric domains}
\begin{figure}[htp]
	\centering
	\includegraphics[width=1\columnwidth]{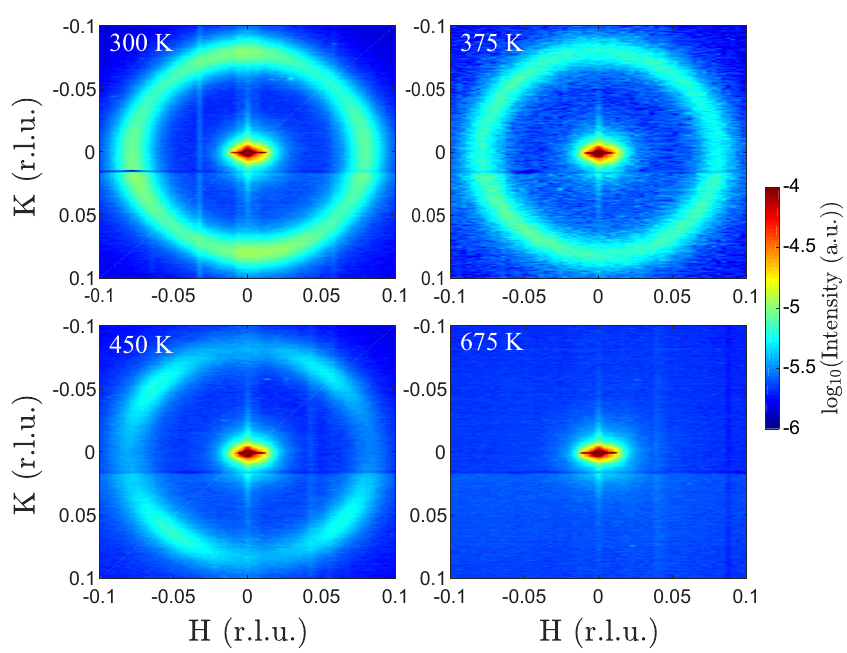}
	\caption{In-plane RSMs around the 002 superlattice Bragg peak. The ring of intensity is due to diffuse scattering from periodic ferroelectric domains, with a period $\Lambda_d \simeq$ 5 nm. The domain structure evolves from walls lying predominantly in the $\left\lbrace 100 \right\rbrace $ planes at room temperature, to a more isotropic configuration at 375 K (uniform ring), to  $\left\lbrace 110 \right\rbrace $ DWs at 450 K. No diffuse scattering is observed in the paraelectric phase at 675 K.}
	\label{fig:fig1}
\end{figure}

Fig. \ref{fig:fig1} shows in-plane reciprocal space maps (RSMs) around the 002 Bragg peak of a (PTO$_5$$\vert$STO$_4$)$_{28}$ superlattice as a function of temperature (here and throughout we label the reflections according to the average perovskite lattice parameter $\sim$4 \textup{\AA}; thus in this case the 002 reflection corresponds to the 0018 superlattice peak). The kink in tetragonality corresponding to the ferroelectric-to-paraelectric phase transition in this superlattice was found to occur at \textup{$T_C$} = 570 K. The in-plane RSMs are plotted in reciprocal lattice units (r.l.u.) of the substrate (1 r.l.u. = 2$\pi$/3.905 $\textup{\AA}^{-1}$) and were obtained by integrating the diffracted intensity along the out-of-plane direction in reciprocal space over the range $\Delta L=\pm0.03$ r.l.u. around the superlattice Bragg peak. The ring of diffuse scattering around the superlattice Bragg peak is due to the ferroelectric domains in PbTiO$_3$ with a periodicity $\Lambda_d$ = 3.905 $\textup{\AA}/\Delta H \simeq$ 5 nm at room temperature \cite{Streiffer2002}. 

At 300 K, a diffuse ring of intensity is observed. The intensity distribution, however, is not uniform around the ring, with the highest intensity observed at points where the ring crosses the $H$ and $K$ axes, indicating that DWs preferentially lie in the $\left\lbrace 100 \right\rbrace$ crystallographic planes. As the temperature is increased to 375 K, the ring becomes more uniform. At still higher temperatures (e.g. 450 K), the dominant intensity in the diffuse scattering is redistributed towards four spots with $H = \pm K$, indicating that DWs lie preferentially in the $\left\lbrace 110 \right\rbrace $ planes. The bottom right figure shows the same RSM in the paraelectric phase, where the polarization vanishes and domains disappear.

\begin{figure}[hbtp]
	\centering
	\includegraphics[width=1\columnwidth]{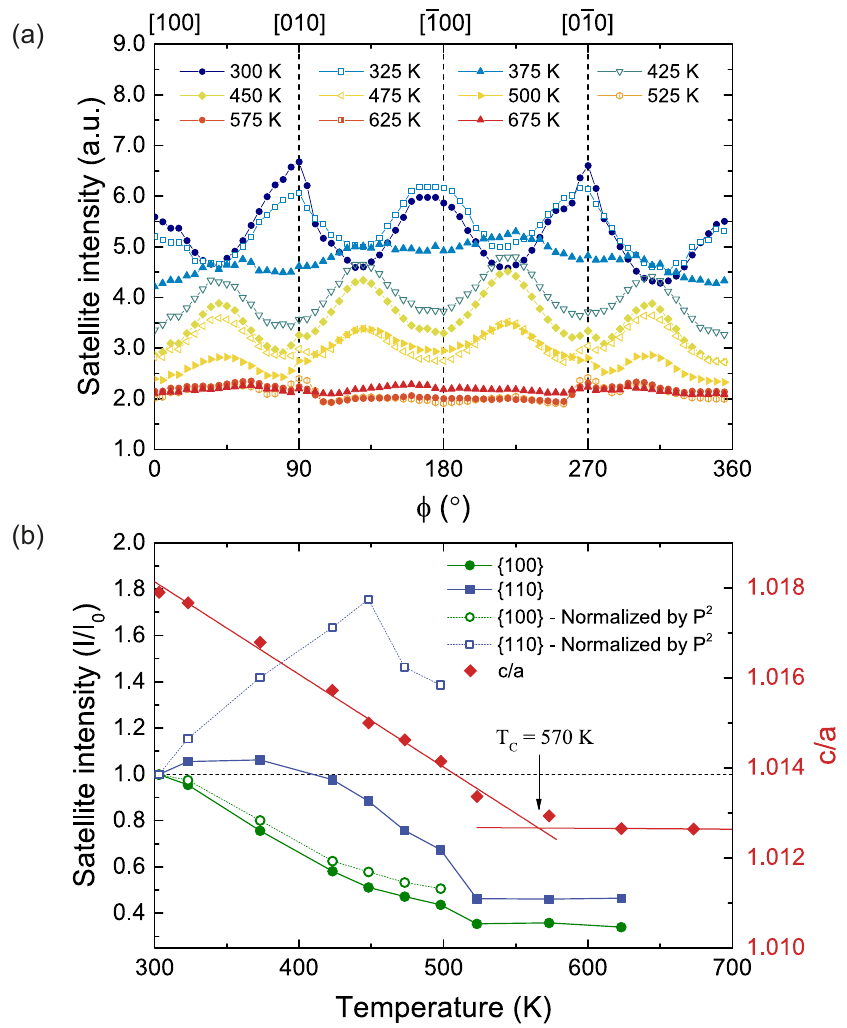}
	\caption{(a) Diffuse scattering intensity as a function of azimuthal angle $\phi = \arctan(K/H)$, plotted at various temperatures. (b) Average tetragonality ($c/a$) and domain satellite intensities for the two DW orientations normalized by the values at room temperature. Hollow markers correspond to intensities first normalized by the square of the polarization.}
	\label{fig:fig2}
\end{figure}

To see the domain reorientation more clearly, we have defined an annulus in reciprocal space (from $|Q|=0.04$ r.l.u. to $|Q|=0.06$ r.l.u., where $|Q|=\left(H^2+K^2\right)^{1/2}$) which only includes the diffuse scattering intensity due to the domains. By looking at the intensity as a function of azimuthal angle $\phi = \arctan(K/H)$ (Fig. \ref{fig:fig2}(a)) we can clearly see that the $\left\lbrace 100 \right\rbrace $ domain walls (DWs) are dominant at room temperature, whereas the $\left\lbrace 110 \right\rbrace $ DWs are dominant above approximately 375 K. 

%By looking at the intensity as a function of azimuthal angle (Supplementary Fig. S1 \cite{Supplemental2017}) we can clearly see that the $\left\lbrace 100 \right\rbrace $ DWs are dominant at room temperature, whereas the $\left\lbrace 110 \right\rbrace $ DWs are dominant above approximately 375 K. 

There are two contributions to the observed intensity changes with increasing temperature: (i) the intensity of the diffuse scattering from the domain structure is reduced due to the reduction of the spontaneous polarization, and (ii) there is a continuous change in the volume fraction of the two orientations as the $\left\lbrace 100 \right\rbrace $ DWs rotate to lie within the $\left\lbrace 110 \right\rbrace $ planes. To separate the first effect we have measured the $c$ and $a$ lattice parameters at each temperature and normalized the domain satellite intensities by the corresponding $\left(c/a\right) - \left(c/a\right)_{para}$ (where $\left(c/a\right)_{para}$ is the tetragonality in the paraelectric phase), which is expected to be proportional to the square of the spontaneous polarization within the domains \cite{Lichtensteiger2005,Dawber2007}. We then find that the normalized intensity of the $\left\lbrace 110 \right\rbrace $ DWs increases with temperature, whereas the $\left\lbrace 100 \right\rbrace $ intensity decreases (Fig. \ref{fig:fig2}(b)). This leads us to conclude that there is a gradual reduction in the volume fraction of $\left\lbrace 100 \right\rbrace $ DWs as they reorient towards the $\left\lbrace 110 \right\rbrace $ planes. This conclusion is corroborated by the observation that the in-plane correlation length of the domain satellites along the $\left\langle 110 \right\rangle $ directions increases with increasing temperature (whereas it decreases along other directions), signalling an increase in order along $\left\langle 110 \right\rangle $.

The same reorientation was also observed for samples with different layer periodicities (not shown). The temperature range of the reorientation was found to depend on the domain period and ferroelectric layer thickness, with $\left\lbrace 110 \right\rbrace $ DWs being more pronounced (observed to be dominant across a wider range of temperatures) for samples with thinner layers of PbTiO$_3$ (and hence smaller domain periods).

We note that these observations differ from those of Fong et al. \cite{Fong2004}, who found that in ultrathin PbTiO$_3$ films, the diffuse scattering due to the periodic domain structure instead evolves from a uniform ring (F$_\beta$ phase) at lower temperatures to a structure with $\left\lbrace 100 \right\rbrace $ DWs (F$_\alpha$ phase) upon heating, accompanied by an abrupt decrease in the domain period. We also observe a gradual decrease in the domain period with temperature, but no abrupt changes, consistent with previous reports \cite{Zubko2012,Boulle2016}. Our observations are also different from the abrupt change in the alignment of ferroelastic DWs observed on heating in BaTiO$_3$ thin films \cite{Everhardt2016}. 

First-principles simulations of PbTiO$_3$/SrTiO$_3$ superlattices strained to SrTiO$_3$ have shown that the two domain orientations are very close in energy \cite{Aguado-Puente2012}, consistent with the above experimental observation of ring-like features in diffuse scattering from domains in superlattices and the meandering DWs in thicker PbTiO$_3$ films observed by scanning probe techniques in Refs. \onlinecite{Thompson2008,Lichtensteiger2014}. Model Hamiltonian calculations predict that increasing compressive strain induces a change from $\left\lbrace 100 \right\rbrace $ to $\left\lbrace 110 \right\rbrace $ DWs concomitant with a structural transition from a phase with non-zero in-plane polarization components to one with purely out-of-plane polarized domains \cite{Jiang2014}. Although the temperature dependence of the DW orientation was not explicitly studied in Ref. \onlinecite{Jiang2014}, since the same phase transition is accessible by increasing temperature, a similar DW reorientation might be expected on heating, as indeed observed experimentally here. The prediction that thinner films favor $\left\lbrace 110 \right\rbrace $ DWs \cite{Jiang2014} is also consistent with our observations. 

Since both thickness reduction and temperature increase are expected to broaden DWs \cite{LukYanchuk2009}, we speculate that the DW width plays an important role in the observed DW reorientation. The multicomponent nature of the order parameter in ferroelectrics also means that changes in the DW character (e.g. the effect of in-plane polarization components) \cite{Aguado-Puente2012,Chapman2017,Wojde??2014,Marton2013,Lee2009}  and chirality are also likely to influence both the DW width \cite{Houchmandzadeh1991} and preferred  orientation; further experimental and theoretical work is needed to clarify this issue.  

\subsection{\label{sec:uniform_region}Real-space mapping of domain orientations with a nanofocused beam}
\begin{figure}[t]
	\centering
	\includegraphics[width=1\columnwidth]{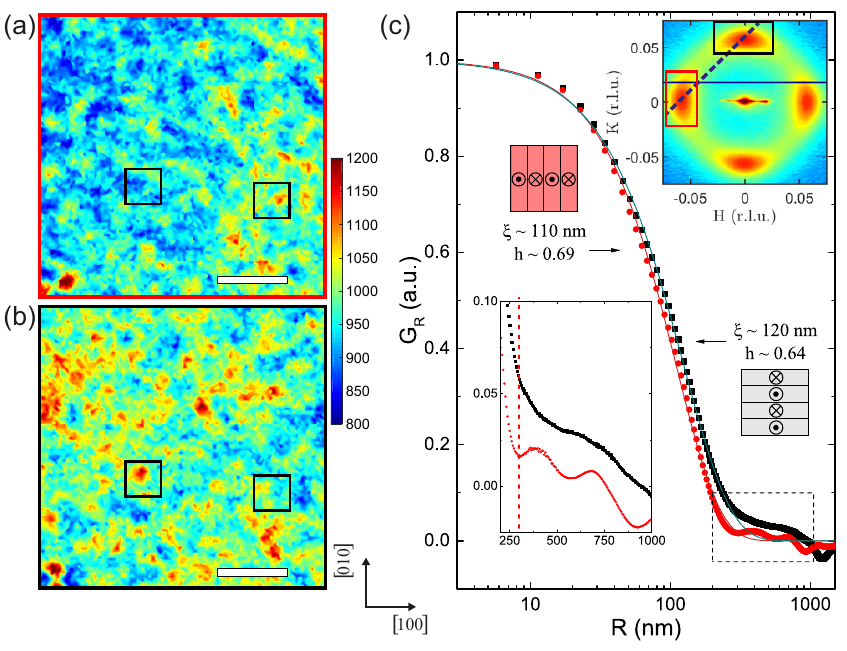}
	\caption{Spatial variation of domain satellite intensity for $\left(010\right)$ (a) and $\left(100\right)$ (b) DWs. Horizontal scale bars correspond to 1 \textmu{}m. The black boxes show two anti-correlated regions. (c) Normalized average autocorrelation functions for images in (a) (red circles) and (b) (black squares). Fitting parameters $\xi$ and $h$ are discussed in the main text. The bottom left inset is an expanded plot around the dashed box in the main figure, showing a minimum at 300 nm for one DW orientation. The top right inset shows the in-plane RSM around the 003  peak of the superlattice (unfocused beam). The dashed blue line indicates the approximate orientation of the detector in reciprocal space, whereas the red and blue boxes indicate the domain satellites used for mapping the intensity in (a) and (b) respectively.}
	\label{fig:fig3}
\end{figure}

In order to gain a more local picture of the domain wall orientations, we use a nanofocused beam to measure the domain satellite intensities for a (PTO$_8$$\vert$STO$_4$)$_{19}$ superlattice (with domain period $\Lambda_d \simeq$ 7 nm) as a function of position on the sample by translating it relative to the beam using a piezo stage. The resulting maps of domain satellite intensity for $\left(010\right)$ and $\left(100\right)$ DWs are shown in Figs. \ref{fig:fig3}(a) and \ref{fig:fig3}(b) respectively. Comparing the two images (see for example the features in the square boxes) reveals that some regions are anticorrelated with maxima in the intensity of $\left(100\right)$ DWs corresponding to minima in the intensity of $\left(010\right)$ and vice-versa. Other regions in the images are not fully anticorrelated, presumably because they correspond to regions with similar fractions of both orientations or with DWs that are oriented along other directions.  

The maximum intensity variation across the images is around 30\%, indicating that both domain orientations are present within the beam footprint at each point and therefore the patchy features in the images correspond to local variations of the predominant DW orientations, rather than individual regions with a specific, well-defined DW orientation. A lower bound on the size of the latter can be determined from the in-plane correlation length obtained from a Scherrer-type analysis of the FWHM of the domain satellites (measured using an unfocused beam), which we estimate to be around 30 nm. This is several times smaller than the footprint of the focused beam and thus below the resolution of the local measurements.

To estimate the characteristic length scale $\xi$ for the features in Figs. \ref{fig:fig3}(a) and \ref{fig:fig3}(b) we calculate the autocorrelation function (ACF) of each image. If $I(\mathbf{x})$ is the satellite intensity at position $\mathbf{x}$, the 2D ACF is defined as $G(\mathbf{R})=\left\langle I(\mathbf{x})I(\mathbf{x}-\mathbf{R}) \right\rangle $, where $\mathbf{R}$ is a displacement vector and averaging is performed over all possible $\mathbf{x}$. For simplicity, we assume an isotropic distribution of regions with a given DW orientation and use the 2D ACF to calculate the azimuthal average over all in-plane directions $G(R)$ with $R=|\mathbf{R}|$. To extract $\xi$, we fit  $G(R)$ to: 
\begin{equation}
\label{eq:acf}
G\left(R\right)=\sigma_I^2\exp\left[ \left(-R/\xi\right)^{2h}\right].
\end{equation}
This expression \cite{Sinha1988} has previously been used to estimate domain sizes measured using piezoresponse force microscopy \cite{Shvartsman2005,Yao2012}. In the above, the function $I(\mathbf{x}) - I(\mathbf{x}-\mathbf{R})$ is assumed to be a Gaussian random variable with a standard deviation $\sigma_I$, and $h$ is a parameter that describes the abruptness of intensity changes ($h$ is small for very abrupt changes, whereas it tends to unity for smoother variations) \cite{Sinha1988}. 

The average ACFs corresponding to the images in Fig. \ref{fig:fig3}(a) and \ref{fig:fig3}(b) (after subtraction of the mean intensity of each image) are shown in Fig. \ref{fig:fig3}(c). Fitting the autocorrelation data, we estimate the characteristic length scale $\xi$ to be of the order of 110-120 nm, which is comparable to the beam footprint. We note that the length scale extracted from Equation \ref{eq:acf} is always a lower estimate of the feature size \cite{Matzler1997} and is usually quoted in cases where there is a distribution of sizes. In images with a well-defined feature size, a better measure of this size is the first minimum in the ACF \cite{Fernandez-Pena2016}. This minimum is discernible for only one orientation (Fig. \ref{fig:fig3}(a)), and occurs at around 300 nm.  

\begin{figure}[t]
	\centering
	\includegraphics[width=1\columnwidth]{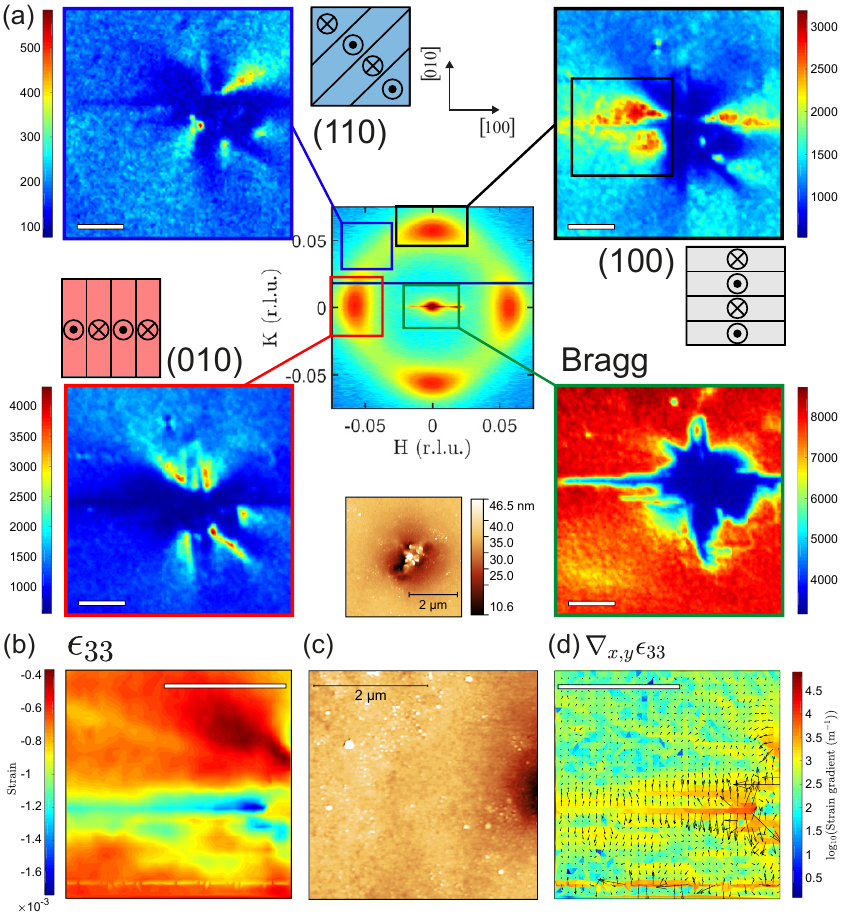}
	\caption{(a) Domain satellite intensity around a defect in the superlattice, showing DW alignment parallel to the long features of the defect. The schematics show the probed DW orientation for each image, with the central RSM indicating the RS regions probed in each map. The bottom right image is a map of the Bragg peak intensity around the same region. The color bars show the intensity in arbitrary units. The central bottom figure shows an AFM scan of the topography at the center of the defect. (b) Map of the strain $\epsilon_{33}$ in the region marked by the box in the top right image of (a). (c) AFM image of the superlattice topography around the same region. (d) Map of the 2D strain gradient $\nabla_{x,y}\epsilon_{33}$ with arrows and colors representing its direction and magnitude respectively. The arrows around the region with the largest strain gradient point to a direction perpendicular to the DWs.  The horizontal scale bars in all images have a size of 2 \textmu{}m.}
	\label{fig:fig4}
\end{figure} 

Performing the same analysis on images from other parts of the sample shows that $\xi$ varies significantly from around 100 nm to microns, and even within individual images, different length scales are apparent on visual inspection. We therefore conclude that although the domain period is well-defined by the electrostatic boundary conditions, as can be seen from the sharpness of the peaks in the RSM of Fig. \ref{fig:fig3}(c), the dominant DW orientation exhibits variations over a wide range of length scales across the sample and the local DW orientation is likely to be determined by other factors, as we discuss next.  

\subsection{\label{sec:etched}Domain wall alignment around topographical defects}

\begin{figure}[h]
	\centering
	\includegraphics[width=1\columnwidth]{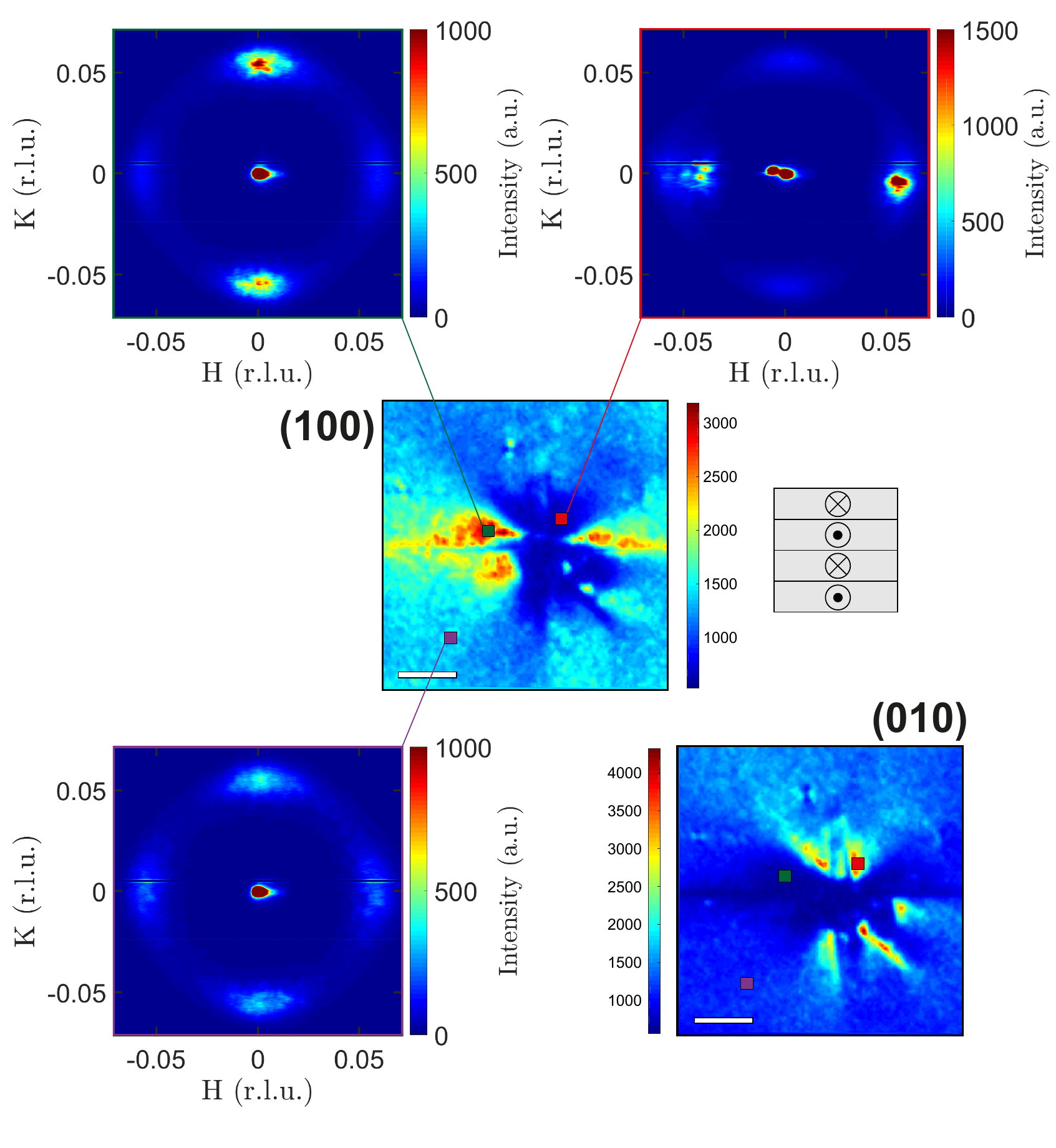}
	\caption{RSMs (linear scale) at different positions on the sample obtained with a nanofocused beam. The central figure shows the spatial variation of the intensity of the domain satellite corresponding to the periodicity along [100]. The top left RSM shows the diffraction ring for a region with preferential (100) domain wall alignment, the top right for a region with preferential (010) alignment and the bottom left for a region with no strong preferential alignment. The bottom right image shows the spatial variation of the intensity of the domain satellite corresponding to the periodicity along [010].}
	\label{fig:fig5}
\end{figure}

To better understand what influences the local DW orientation, we look at a region of the film close to a topographical defect. The defect has an irregular shape, as seen from the AFM image in Fig. \ref{fig:fig4}(a) and gives rise to long, streaky features of reduced intensity in the map of the superlattice Bragg peak intensity (Fig. \ref{fig:fig4}(a), bottom right). Intensity maps of the domain satellites reveal that the domain stripes align preferentially along the length of these features. The intensity in these aligned regions is more than double that in the more uniform parts of the film, showing a strong local preference for that particular DW orientation compared to the rest of the superlattice. Local in-plane RSMs, obtained with the nanofocused beam and shown in Fig. \ref{fig:fig5}, further highlight this preferential alignment. Fig. \ref{fig:fig5} shows three RSMs for: (i) a uniform region with no strong preferential alignment (bottom left), (ii) a region with preferential (100) domain wall alignment (top left) and (iii) a region with preferential (010) alignment (top right). The splitting of the Bragg peak in the bottom right RSM is due to a slight drift of the beam along the rocking curve direction.   

\begin{figure}[hbtp]
	\centering
	\includegraphics[width=1\columnwidth]{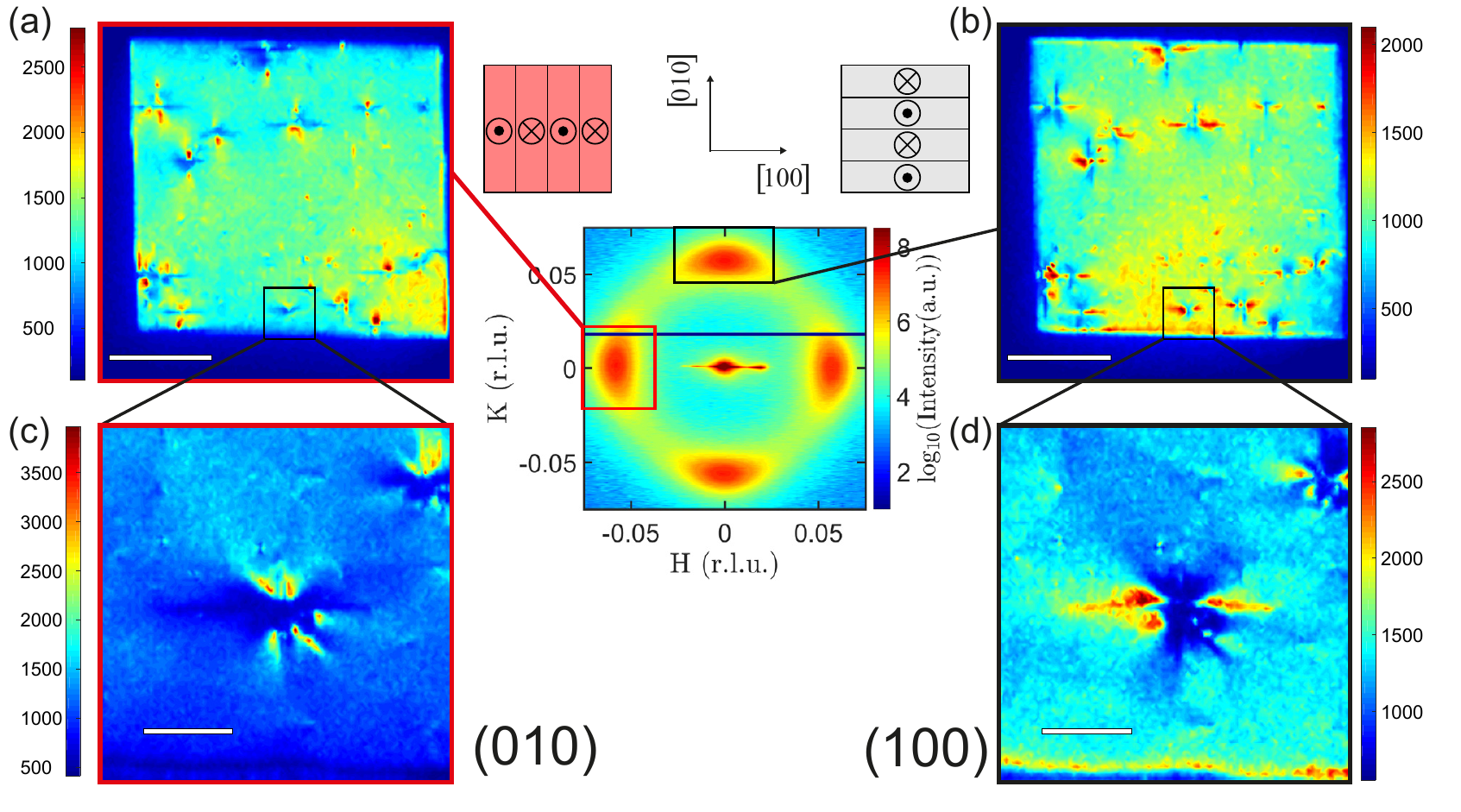}
	\caption{Spatial variation of the domain satellite intensity around the Ar-ion milled capacitor structures for (a) $\left(010\right)$ and (b) $\left(100\right)$ DWs. Scale bars correspond to 40 \textmu{}m. (c) and (d) are smaller scans around the edge of the capacitor, corresponding to scans (a) and (b) respectively, with scale bars corresponding to 5 \textmu{}m. The central figure shows the reciprocal space map around the 003 Bragg peak obtained using a full (approximately 50 \textmu{}m $\times$ 50 \textmu{}m) beam.}
	\label{fig:fig6}
	
\end{figure}

We also observe some preferential DW alignment with the Ar-ion milled edges of the square electrodes. Fig. \ref{fig:fig6} shows the spatial variation of the domain satellite intensity across the capacitor defined by Ar-ion milling. Outside the 120 \textmu{}m $\times$ 120 \textmu{}m capacitor area, the surrounding superlattice was milled away entirely down to the bottom electrode resulting in vanishing intensity of the domain satellites. Figs. \ref{fig:fig6}(a) and \ref{fig:fig6}(b) show the intensity variation of satellites corresponding to (010) and (100) DWs respectively across the sample. Figs. \ref{fig:fig6}(c) and (d) are higher magnification maps of the same satellite intensities near one of the capacitor edges. A preferential alignment of DWs along the edge of the Ar-ion milled capacitor is observed, as is clear from the higher intensity of the corresponding domain satellite at the bottom of Fig. \ref{fig:fig6}(d).

Furthermore, by obtaining a three-dimensional RSM at each point, we can map the variation of the out-of-plane lattice parameter of the film around this region and extract the local distribution of the average out-of-plane strain $\epsilon_{33}$ that accompanies the DW alignment. The corresponding map of $\epsilon_{33}$ is shown in Fig. \ref{fig:fig4}(b), together with an AFM scan of the local superlattice topography (Fig. \ref{fig:fig4}(c)). We can further use the strain map to extract the strain gradients $\nabla_{xy}\epsilon_{33}$ ($\nabla_{xy}=\mathbf{i}\frac{\partial}{\partial x}+\mathbf{j}\frac{\partial}{\partial y}$) around this region (Fig. \ref{fig:fig4}(d)). We find that the regions with preferential domain wall alignment are correlated with changes in strain of the order of 0.1\% and strain gradients of the order of 10$^4$ m$^{-1}$ directed perpendicular to the DWs, as shown by the arrows in Fig. \ref{fig:fig4}(d). 

Preliminary studies of the temperature dependence also reveal that the aligned DWs around the defect return to the same configuration when the superlattice is heated above its transition temperature and subsequently cooled to room temperature. The preferential DW alignments reported here are in line with previous observations that DWs in ultrathin PbTiO$_3$ films align with the crystallographic steps of the substrate \cite{Streiffer2002,Fong2004,Thompson2008,Prosandeev2007}. The precise mechanism for this alignment remains to be fully understood, but our findings suggest that by nanopatterning the films, conventional lithography methods could be used to define local DW orientation in ferroelectric thin films. 

%\section{\label{sec:conclusions}Conclusions}
To conclude, we have studied the behavior of ferroelectric domains in PbTiO$_3$/SrTiO$_3$ superlattices with synchrotron X-ray diffraction. We discover a temperature regime where the DW normals rotate from preferential $\left\langle 100 \right\rangle $ alignment at low temperatures to $\left\langle 110 \right\rangle $ at high temperatures. By probing the spatial distributions of the domain orientations using a nanofocused beam we were able to map the local orientations of buried domains in these superlattices. We observe that in the vicinity of a structural defect the DWs are preferentially aligned along specific directions, and map out the inhomogeneous strain and strain gradient fields associated with this defect. A similar preferential DW alignment is observed along the ion-milled edges of the film, offering a route to controlling the behavior and orientation of domains in ferroelectric thin films that may help further advance the growing field of DW nanoelectronics.  

This work was supported by the EPSRC (Grant numbers EP/M007073/1, P.Z. and 1447654, M.H.), the A.G. Leventis Foundation (M.H.), and the UCL-ESRF Impact studentship scheme (E.Z.). We acknowledge the European Synchrotron Radiation Facility and the ID01 beamline staff for support during the experiment.

\end{document}